\documentclass[aps,prl,preprint,groupedaddress]{revtex4}
\usepackage{graphicx}

\begin{document}


\title{Criticality of Electron-Nucleus Cusp Condition to Local Effective Potential Energy Theories}


\author{Xiao-Yin Pan and Viraht Sahni }
\affiliation{Department of Physics, Brooklyn College of the City
University of New York, 2900 Bedford Avenue, Brooklyn, New York
11210, and The Graduate School of the City University of New
York,360 Fifth Avenue, New York, New York 10016. }


\date{\today}

\begin{abstract}
  Local(multiplicative) effective potential energy theories of electronic structure
  comprise the transformation of the Schr{\"o}dinger equation  for
  interacting fermi systems to model noninteracting fermi or bose
  systems whereby the equivalent density and energy are obtained.
  By employing the   integrated
  form of the Kato electron-nucleus cusp condition, we prove that the effective electron
  -interaction potential energy of these model fermions or bosons is finite at a
  nucleus. The proof is general and valid for arbitrary system whether it be
  atomic, molecular, or solid state, and for arbitrary state and symmetry.
  This then provides justification for all prior work in the
  literature based on the assumption of finiteness of this
  potential energy at a nucleus. We further demonstrate the
  criticality of the electron-nucleus cusp condition to such
  theories by example of the Hydrogen molecule. We show thereby
  that both model system effective electron-interaction potential
  energies, as determined from densities derived from accurate
  wave functions, will be singular at the nucleus unless the wave
  function satisfies the electron-nucleus cusp condition.

\end{abstract}


\maketitle

\section{I. Introduction}
    Consider the time-independent Schr{\"o}dinger equation for atoms, molecules, or solids:
    \begin{equation}
    {\hat H} \Psi ({\bf r}_{1},...{\bf r}_{N})= E \Psi ({\bf r}_{1},...{\bf r}_{N}),
    \end{equation}
 where $  \Psi ({\bf r}_{1},...{\bf r}_{N})$ is the system wave function, $E$ the energy,
 and $N$ the number of electrons. (We suppress the spin coordinate throughout the paper.)
  The electronic Hamiltonian ${\hat H}$ in atomic units is
    \begin{equation}
    \hat {H} =- \frac{1}{2}\sum_{i} \nabla_{i}^{2} + \sum_{i}v({\bf r}_{i})+ \frac{1}{2}\sum_{i,j}^{\prime} \frac{1}{\left|{\bf r}_{i}-{\bf r}_{j}\right|},
      \end{equation}
  where the first and last terms represent the electron kinetic and electron interaction
  potential energy operators, respectively, and $v({\bf r})$ the operator representing
  the potential energy of the electrons due to the external charge $Z$ of the nuclei at ${\bf R}_{\alpha}$:
  \begin{equation}
  v({\bf r})=\sum_{\alpha}\frac{Z}{\left|{\bf r}-{\bf R}_{\alpha}\right|}.
   \end{equation}
 The electronic density $\rho({\bf r})$ of the system is defined as
    \begin{equation}
  \rho({\bf r})=N \int \psi^{*}({\bf r},{\bf r}_{2}...{\bf r}_{N}) \psi({\bf r},{\bf r}_{2}...{\bf r}_{N}) d{\bf r}_{2} ... d{\bf r}_{N}.
    \end{equation}
At the coalescence of an electron with a nucleus, the external
potential energy operator
  $v({\bf r})$ is singular. For the wave function to satisfy the Schr{\"o}dinger equation and remain bounded, it must satisfy the Kato \cite{1} electron nucleus cusp condition at each
  nucleus. The cusp condition is usually stated in differential form \cite{1}.
  In integrated form, the statement of the cusp condition for arbitrary state of the system is \cite{2,3}
    \begin{equation}
  \psi({\bf r},{\bf r}_{2}, ...{\bf r}_{N})=\psi(0,{\bf r}_{2}, ...{\bf r}_{N}).(1-Z r)+
   {\bf r}\cdot {\bf a}({\bf r}_{2}, ...{\bf r}_{N})+ O(r^{2}),
  \end{equation}
  where the vectors ${\bf r}_{i}$ now represent the positions of the electrons from a particular nucleus.
  The vector  ${\bf a}({\bf r}_{2}, ...{\bf r}_{N})$ is undetermined.
 The derivation of the integrated form of the cusp condition does not involve any
          boundary conditions far from the nucleus.  Hence, it is valid at the nuclei of atoms,
          molecules, and periodic and aperiodic solids.\\

     In addition to solving the Schr{\"o}dinger equation(1), electronic structure is also
      determined by calculations performed within the framework of local(multiplicative)
   effective potential energy theories. Representative of such theories are Kohn-Sham density functional theory (KS-DFT) \cite{4}, the Optimized Potential Method (OPM) \cite{5},
   and quantal density functional theory (Q-DFT) \cite{6}. The basic idea underlying such theories is the transformation of the Schr{\"o}dinger equation to that of the model S system of \textit{noninteracting fermions} whereby the equivalent density $\rho ({\bf r})$
    and energy $E$ are obtained.  The differential equation describing the model fermions is
    \begin{equation}
   [-\frac{1}{2}\nabla^{2}+ v_{s}({\bf r})]\phi_{i}({\bf r})=\epsilon_{i} \phi_{i}({\bf r});  \; i=1,2,..N,
    \end{equation}
    where $v_{s}({\bf r})$ is their effective potential energy:
       \begin{equation}
      v_{s}({\bf r})=v({\bf r})+v_{ee}({\bf r}),
    \end{equation}
     with $v_{ee}({\bf r})$ their effective electron-interaction potential energy.
        The potential energy $v_{ee}({\bf r})$ accounts for electron correlations due to the Pauli exclusion principle and Coulomb repulsion, and Correlation-Kinetic effects which are a
      consequence of the difference in the kinetic energy of the interacting and noninteracting  systems. The various theories differ in their definitions of the potential
      energy $v_{ee}({\bf r})$. In KS-DFT, the energy $E$ is a functional of the ground state density
    $\rho ({\bf r})$, and   $v_{ee}({\bf r})$ is defined as the functional derivative
    of its KS electron-interaction  energy $E^{KS}_{ee}[\rho ]$ component. In the OPM, this
    energy component is considered a functional of the orbitals $\phi_{i}({\bf r}): E^{KS}_{ee}=  E^{KS}_{ee}[\phi_{i}]$. The potential energy $v_{ee}({\bf r})$, defined as
    the functional derivative of $E^{KS}_{ee}[\phi_{i}]$, is obtained by solution of an
    integral equation in conjunction with the differential equation (6). Within Q-DFT,
    the potential energy $v_{ee}({\bf r})$ is the work done to move the model fermion in a
    conservative field ${\cal F}_{s}({\bf r})$. The density $ \rho({\bf r})$ equivalent to that of the interacting system  Eq. (4)
    is determined from the orbitals $\phi_{i}({\bf r})$ as
     \begin{equation}
     \rho({\bf r})=\sum_{i} \left| \phi_{i}({\bf r})\right|^{2}.
     \end{equation}
   The energy is determined from the energy functionals, or in terms of the components of
   the conservative field. The highest occupied eigenvalue of the differential  equation
   is the negative of the
   ionization energy \cite{7,8,9}.\\

An understanding of the structure and general properties of the
potential energy $v_{ee}({\bf r})$ that allows for the
transformation from the interacting to the noninteracting model
system is therefore of importance. The structure is also of
significance for the evaluation and construction of approximations
within the various S system theories.  A key aspect of this
structure, and one that has been controversial, is whether the
potential energy $v_{ee}({\bf r})$ is finite or singular at a
nucleus. The principal manner by which the structure is determined
is one that employs methods\cite{10} which assume knowledge of the
'exact' density $ \rho({\bf r})$. The densities in turn are
obtained from correlated or configuration-interaction wave
functions that are highly accurate from the total energy
perspective.  Work on the He atom by Smith et al \cite {11},
Davidson \cite{12} , and Umrigar and Gonze \cite{13}, show
$v_{ee}({\bf r})$ to be finite at the nucleus. Almbladh and
Pedroza \cite{14}, on the other hand, showed it to be singular
there. Additional work on light atoms show it to be either finite
\cite{13,15} or to diverge at the nucleus \cite{16}. For the
determination of its structure for few- electron molecular systems
\cite{17} such as $H_{2}$ and $LiH$, the potential energy
$v_{ee}({\bf r})$ is assumed finite at each nucleus. Expressions
for $v_{ee}({\bf r})$ at a nucleus have also been derived
\cite{18}, but once again they are based on the assumption that it
is finite there. In various approximations with KS-DFT \cite{19},
 the potential energy $v_{ee}({\bf r})$ also diverges at the
 nucleus.\\

The controversy was resolved for closed shell atoms and open shell
atoms in the central field approximation by Qian and Sahni
\cite{20} who proved analytically that $v_{ee}({\bf r})$ is in
fact finite at the nucleus. Furthermore,  they were the first to
show  that this finiteness was a direct consequence of the
electron-nucleus cusp condition.  In their proof they employed the
differential form of the cusp condition. In this form of the cusp
condition, the angular dependence  is integrated out. Hence, its
application constrains their proof  to spherically
symmetric systems.\\

In this paper we generalize the conclusion of Qian and Sahni and
prove that $v_{ee}({\bf r})$ is finite at a nucleus independent of
the type of system (atomic, molecular, or solid state), and of the
system state and symmetry. Our proof employs instead the
integrated form of the cusp condition, and it is for this reason
that the result is valid for systems of arbitrary symmetry. The
proof too is distinctly different. \emph{Ex post facto}, the proof
thus provides justification for all the work on the determination
of $v_{ee}({\bf r})$ based on the assumption that it is finite at a nucleus.\\

  It is also possible to transform the Schr{\"o}dinger equation to that of the B system of \textit{noninteracting bosons }  such that the density $\rho ({\bf r})$ and energy
  $E$ of the interacting system is once again obtained \cite{8}. In this local effective potential
  energy theory, the density amplitude $\sqrt{\rho ({\bf r})}$ is determined directly.
   The differential equation for the model bosons is
     \begin{equation}
   [-\frac{1}{2}\nabla^{2}+ v_{B}({\bf r})]\sqrt{\rho ({\bf r})}=\mu \sqrt{\rho ({\bf r})},
    \end{equation}
  where $v_{B}({\bf r})$ is their effective potential energy:
   \begin{equation}
      v_{B}({\bf r})=v({\bf r})+v_{ee}^{B}({\bf r}),
    \end{equation}
   with  $ v_{ee}^{B}({\bf r})$ the corresponding effective electron-interacting potential
   energy. The potential energy $v_{ee}^{B}({\bf r}) $ accounts for Pauli and Coulomb
    correlations, and Correlation-Kinetic effects due to the difference in kinetic
    energy of the
   interacting fermion and noninteracting  boson systems\cite{21}.  Once again, within
   density functional theory, $v_{ee}^{B}({\bf r})$ is defined \cite{8} as a functional derivative
    of an electron-interaction energy functional $E^{B}_{ee}[\rho]$,
   whereas in Q-DFT \cite{21} it is the work done in a conservative field
    $\cal {F}_{B}({\bf r})$.
   The energy is determined from the total energy functional or in
   terms of the components of the conservative field. The single eigenvalue $\mu$ is
   the chemical potential or the negative of the ionization energy.
    \\

   In this paper we also prove that the potential energy  $v^{B}_{ee}({\bf r})$
   is finite at a nucleus.  The proof is again general and valid for
   systems of \textit{arbitrary} state and symmetry, and also employs the integrated form
   of the cusp condition. \\

   As a second component to the paper, we demonstrate
   the criticality of the electron-nucleus cusp condition to the finiteness of
   the potential energies $v_{ee}({\bf r})$ and $v^{B}_{ee}({\bf r})$ at a nucleus  by application
   to the Hydrogen molecule. It becomes evident thereby that densities derived
   from wave functions that do not satisfy the cusp condition lead to potential energies
   that are singular at a nucleus, irrespective of how accurate the wave functions may be
   from an energy standpoint.\\

\section {II. Criticality of Cusp Condition---Application to the Hydrogen Molecule}
   For two electron systems in their ground state such as the Helium atom, Hooke's atom,
   and Hydrogen molecule, the S and B systems are equivalent. This is because the S system
   orbital is then $\phi_{i}({\bf r})=\sqrt{\rho({\bf r})/2}, i=1,2.$ Hence, the demonstration of the significance of the cusp condition as applied to the Hydrogen
   molecule is equally valid for both systems. Inverting equations (9) or (6), we then have
   for the Hydrogen molecule
   \begin{equation}
   v_{ee}({\bf r})=v_{ee}^{B}({\bf r})=\mu + \frac{\nabla ^{2}\sqrt{\rho({\bf r})}}{2 \sqrt{\rho({\bf r})}}-v({\bf r}).
   \end{equation}
   (In this example, the S system differential equation has only one eigenvalue.)
   It is evident, therefore, that the singularity in $v({\bf r})$ at each nucleus must
   be cancelled by the $  \frac{\nabla ^{2}\sqrt{\rho({\bf r})}}{2 \sqrt{\rho({\bf r})}}$
   term in order for $v_{ee}({\bf r})$ or $v_{ee}^{B}({\bf r})$ to be finite there.\\

   In our calculations we employ the accurate gaussian geminal wave function of Komasa and
   Thakkar \cite{22} for the Hydrogen molecule which is spin free and of the form
   \begin{equation}
    \psi({\bf r}_{1}, {\bf r}_{2})=\frac{1}{4}(1+{\hat P}_{12}) (1+ {\hat P}_{ab}) \sum_{k=1}^{150} c_{k} \phi_{k}
      \end{equation}
   in which
   \begin{equation}
   \phi_{k}=exp(-\alpha_{k}\left|{\bf r}_{1}-{\bf R}_{a}\right|^{2}-\beta_{k}\left|{\bf r}_{1}-{\bf R}_{b}\right|^{2}-
   \zeta_{k}\left|{\bf r}_{2}-{\bf R}_{a}\right|^{2}-\eta_{k}\left|{\bf r}_{2}-{\bf R}_{a}\right|^{2}-\gamma_{k}\left|{\bf r}_{1}-{\bf r}_{2}\right|^{2}),
   \end{equation}
   where ${\bf r}_{j}$ for $j\in\{1,2\}$ are the position vector of the electrons, ${\bf R}_{j}$ for $j\in\{a,b\}$
   are the position vectors of the nuclei, ${\hat P}_{12}$ and ${\hat P}_{ab}$ are permutation operators that interchange
   the electronic and nuclear coordinates, respectively, and $c_{k}, \alpha_{k}, \beta_{k}, \zeta_{k}, \eta_{k}, \gamma_{k}$ are variationally determined parameters subject to the square-integrability constraint
    \begin{equation}
    (\alpha_{k}+ \beta_{k})(\eta_{k}+ \zeta_{k})+ \gamma_{k} ( \alpha_{k}+ \beta_{k}+\eta_{k}+ \zeta_{k})> 0
    \end{equation}
   for each k. The individual exponential parameters are allowed to become negative as long as square-integrability
   is satisfied. The ground state energy obtained with this wave function is $E=-1.174475313$
   a. u. and that of the most accurate correlated wave function \cite{23} is $E=-1.174475668$
   a.u. Thus the energy obtained by the gaussian wave function is accurate to the sixth
   decimal place. It is well known that such gaussian geminal or orbital wave functions do
   not satisfy the electron-nucleus cusp condition, and it is for this reason we employ
   this wave function. Additionally, in contrast to wave functions such as the Kolos-Roothan
   type wave function \cite{23,24}, the calculations are analytical.  \\

   In Fig. 1 we plot the density along the nuclear bond axis with the two nuclei on the z
   axis at $R=\pm0.7 a.u.$. It is evident that the density is very accurate right up to the
   nucleus, and on the scale of the figure appears to possess a cusp at each nuclear position.
    However, in magnifying the scale as in Fig.2, We see
   that there is no cusp as expected, and that the density  is smooth across the
   nucleus.\\

    In Fig.3 we plot $v({\bf r})$ and $ \frac{\nabla ^{2}\sqrt{\rho({\bf r})}}{2 \sqrt{\rho({\bf r})}}$ along the z axis about one nucleus. Whereas $v({\bf r})$ is
    singular as expected, the $  \frac{\nabla ^{2}\sqrt{\rho({\bf r})}}{2 \sqrt{\rho({\bf r})}}$ term is finite at the nucleus. Therefore, the singularity in $v({\bf r})$ is not cancelled. Hence, although the wave function is very accurate from the perspective of
    the ground state energy, the fact that it does not satisfy the electron-nucleus
    cusp condition leads to the potential energy $v_{ee}({\bf r})$ (or $v_{ee}^{B}({\bf r})$ )  being singular at each nucleus. In determining these potential energies from accurate
    densities , it is therefore imperative that the densities be obtained from wave functions that satisfy the electron-nucleus cusp condition.
 \begin{figure}
\begin{center}
 \includegraphics[bb=30 90 542 683, angle=90, scale=0.7]{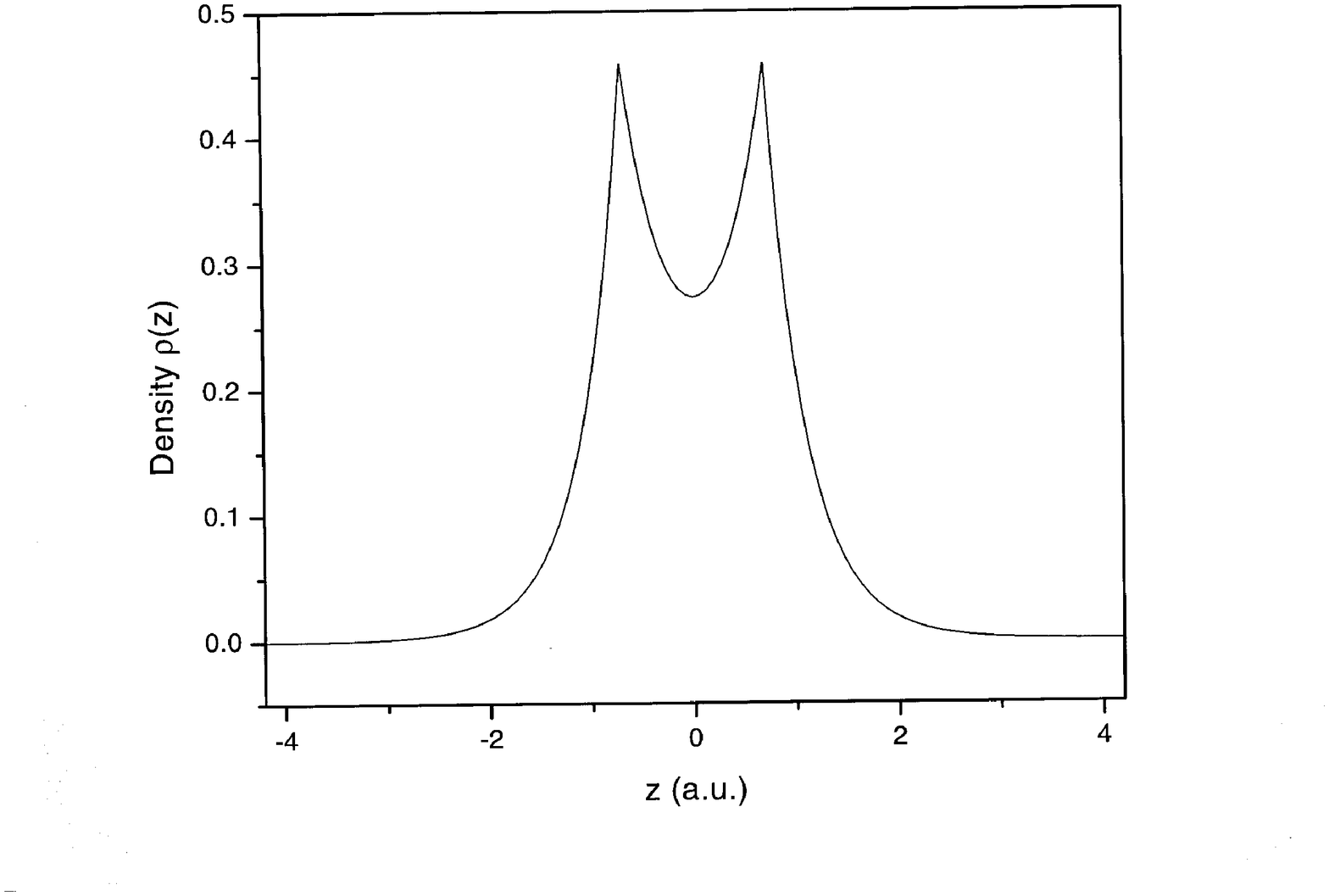}
 \caption{The electron density $\rho({\bf r})$ of the Hydrogen molecule  along
  the nuclear bond axis in atomic units(a.u.). The nuclei are on this axis
  at $\pm 0.7 a.u.$. The density is determined by the
 wave function of Eq.(12).  \label{}}
 \end{center}
 \end{figure}
\begin{figure}
 \begin{center}
 \includegraphics[bb=19 24 590 780, angle=-90, scale=0.7]{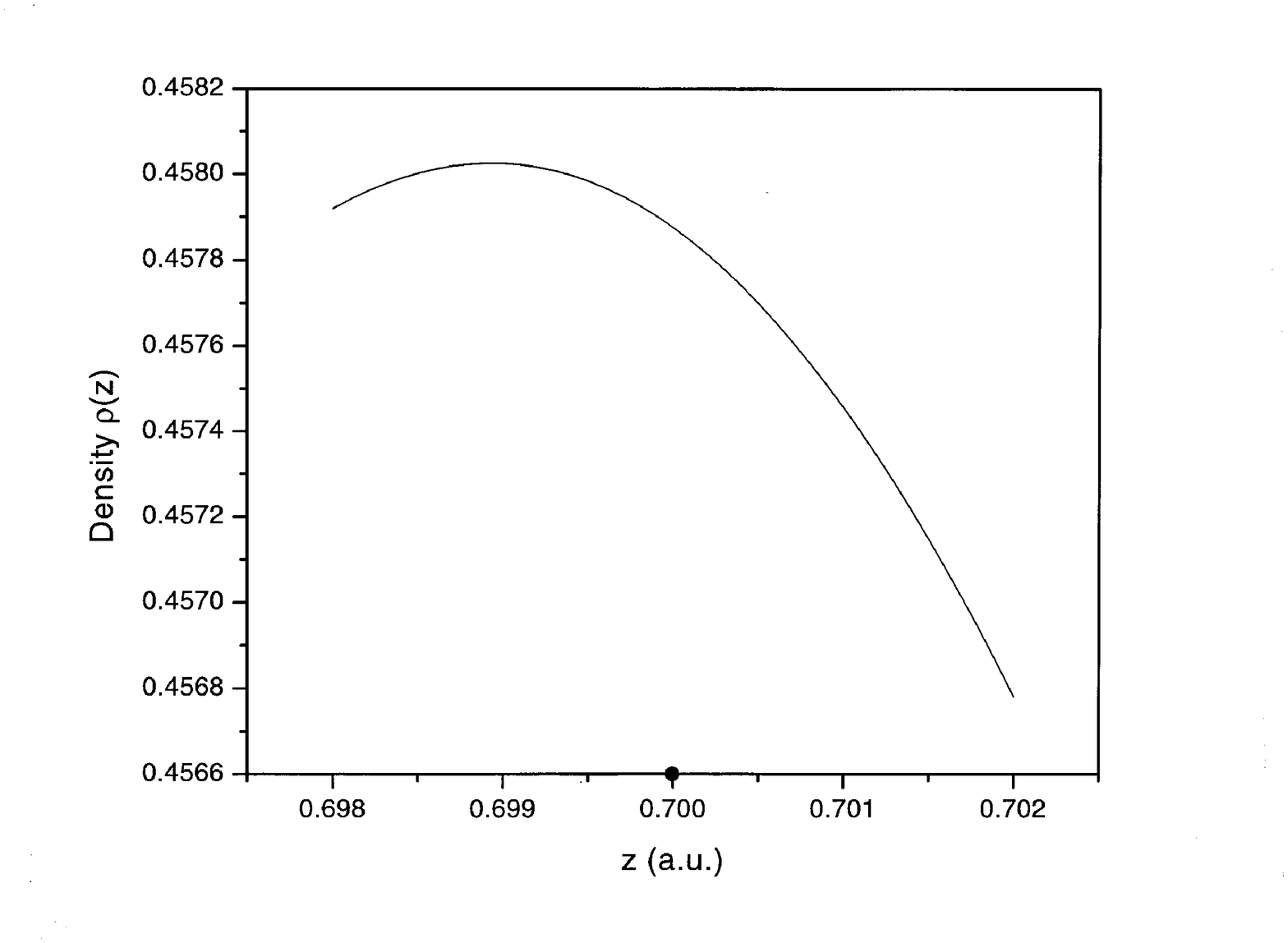}
 \caption{The electron density $\rho({\bf r})$ of the Hydrogen molecule near a
 nucleus as  determined by the wave function of Eq.(12) in atomic units(a.u.). The nucleus is indicated
  by the large dot on the axis.\label{}}
 \end{center}
 \end{figure}
\begin{figure}
 \begin{center}
 \includegraphics[scale=0.7]{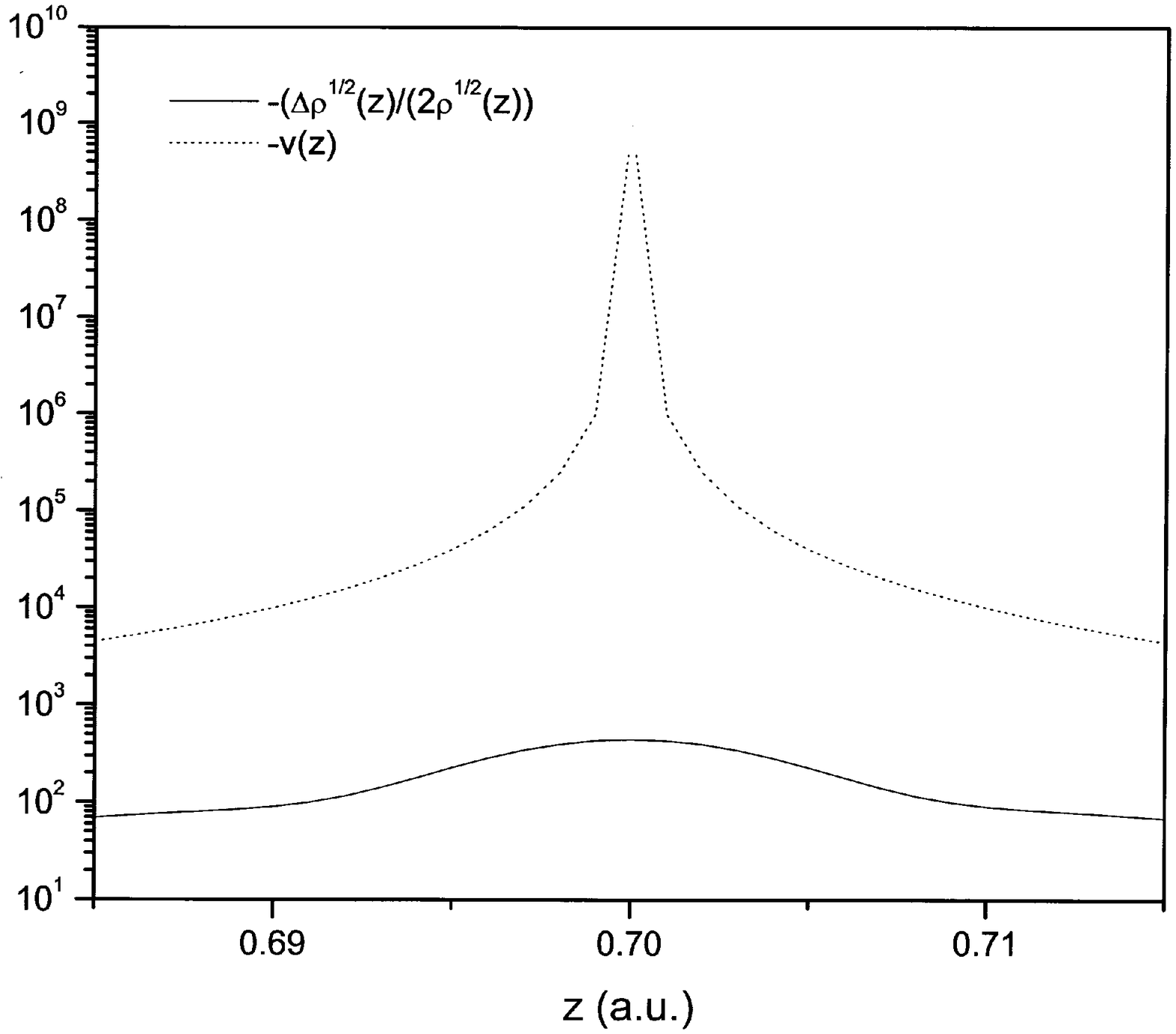}
 \caption{The external potential energy $v({\bf r})$ and the function
$\frac{\nabla^{2} \sqrt{\rho({\bf r})} }{2 \sqrt {\rho({\bf r})}}$
about a nucleus of the Hydrogen molecule in atomic units(a.u.).
The density $\rho({\bf r})$ is determined by the wave function of
Eq. (12). The singularity of the external potential energy at the
nucleus is not cancelled by the  $\frac{\nabla^{2} \sqrt{\rho({\bf
r})} }{2 \sqrt{\rho({\bf r})}}$ function. \label{}}
 \end{center}
 \end{figure}

\section{III . PROOF}
   We first  prove that the potential energy $v_{ee}({\bf r})$ is
   finite at a nucleus.
   Employing the integrated form of the electron-nucleus cusp condition for the
   wave function Eq. (5), the electron density near a nucleus as obtained from the definition of Eq. (4) is
   \begin{equation}
   \rho( {\bf r}) =\rho(0)[(1-Z r)^{2}+2 (1-Z r)\sum _{k=1}^{3} {B}_{k} {r}_{k} +\sum _{l=1,m=1}^{3} r_{l} r_{m} A_{lm}+...],
   \end{equation}
   where $B_{k}= \int a_{k}  \psi(0,{\bf r}_{2}, ...{\bf r}_{N})^{-1}
   d{\bf r}_{2}...d{\bf r}_{N}$ and  $A_{lm}=\int a_{l} a_{m}
   \psi(0,{\bf r}_{2}, ...{\bf r}_{N})^{-2}d{\bf r}_{2}  ...d{\bf r}_{N}$ are constants,
    and $r_{k}, a_{k}$   etc., components of the vector ${\bf r}$ and ${\bf a}$. It follows then that
\begin{eqnarray}
 \sqrt{\rho( {\bf r})}&=&\sqrt{\rho(0)}[(1-Z r)^{2}+2 (1-Z r)\sum _{k=1}^{3} {B}_{k} {r}_{k} +\sum _{l=1,m=1}^{3} r_{l} r_{m} A_{lm}+ ...]^{1/2} \nonumber \\
 &=& \sqrt{\rho(0)}[1-Z r+ {\bf B} \cdot {\bf r}+O(r^{2})],
\end{eqnarray}
   where in the second step we have retained only terms of $O(r)$. \\

   Inverting the S system differential equation we obtain for any occupied orbital
   $\phi_{i}({\bf r})$ the expression for the electron-interaction potential energy
   $v_{ee}({\bf r})$ as
   \begin{equation}
       v_{ee}({\bf r})=\epsilon_{i}+ \frac{\nabla^{2}\phi_{i}({\bf r}) }{2 \phi_{i}({\bf r})}-v({\bf r}).
    \end{equation}
Next, we rewrite the orbitals $\phi_{i}({\bf r})$ as
   \begin{equation}
     \phi_{i}({\bf r})=\sqrt{ \rho({\bf r})} c_{i}({\bf r}), \; i=1,2,...N,
     \end{equation}
    where the coefficients  $c_{i}({\bf r})$ satisfy
     \begin{equation}
     \sum_{i=1}^{N} c_{i}({\bf r})^{2}=1.
     \end{equation}
   This definition of the $\phi_{i}({\bf r})$ is consistent with Eq. (8). Expanding the
   coefficient $c_{i}({\bf r})$ about the nucleus we obtain
   \begin{eqnarray}
    c_{i}({\bf r}) &=&c_{i}(0)+\nabla c_{i}(0) \cdot {\bf r}+O(r^{2})\nonumber \\
    & = & c_{i}(0)[1+{\bf D}\cdot {\bf r}+O(r^{2})],
     \end{eqnarray}
    where ${\bf D}=\nabla c_{i}(0)/c_{i}(0)$ is some constant vector. Inserting Eq. (16) and (20) into Eq. (18), we obtain the expression for the orbitals $\phi_{i}({\bf r})$ near
    the nucleus as
    \begin{equation}
     \phi_{i}({\bf r})=\sqrt{\rho(0)} c_{i}(0)[1-Z r+({\bf B}+{\bf D})\cdot {\bf r}+ O(r^{2})].
     \end{equation}\\

    Now the expression for the external potential energy near the nucleus is
    \begin{equation}
    v({\bf r})=-\frac{Z}{r}-\textstyle{\sum_{\alpha}^{\prime}}\frac{Z}{\left|{\bf r}-{\bf R}_{\alpha}\right|} ,
     \end{equation}
    where the sum is over all the other nuclei. At the nucleus, the term $-\frac{Z}{r}$
    is singular, whereas the other terms are constants. From Eq.(17) it is evident that
    this singularity must be cancelled by the $ \frac{\nabla^{2}\phi_{i}({\bf r}) }{2 \phi_{i}({\bf r})}$ term.\\

    Consider the term $ \frac{\nabla^{2}\phi_{i}({\bf r}) }{2 \phi_{i}({\bf r})}$ near
    the nucleus with $\phi_{i}({\bf r})$ given by Eq. (21). We have $ \nabla^{2} r=2/r$ and
      $\nabla^{2} ({\bf B}+{\bf D})\cdot {\bf r}=0$. After acting by $ \nabla^{2}$ and taking the limit
    as $r\rightarrow 0$, terms of $O(r^{2})$ lead to constants while higher order terms vanish. Thus, near the nucleus, the term $ \frac{\nabla^{2}\phi_{i}({\bf r}) }{2 \phi_{i}({\bf r})}$ is $-Z/r$ plus some constant, and therefore in this limit the singularity
    of the external potential energy is cancelled. Therefore $v_{ee}(0)$ is finite.\\

    The proof of the finiteness of the B system potential energy $v_{ee}^{B}({\bf r})$
at the nucleus is along the same lines as above. Substitution of Eq.(16) into the expression for $v_{ee}^{B}({\bf r})$ of Eq.(11) leads to the result that $v_{ee}^{B}(0)$ is finite.
This result may also be arrived at as a special case of the S system proof for which $c_{i}=1/\sqrt{N}$.\\

  \section{IV. CONCLUSION}
     We have proved that the effective electron-interaction potential energy
     of model \emph{noninteracting} fermi and bose systems that reproduce the density and energy
     of an \emph{interacting} fermi system in an external field  is finite at a nucleus.
     The proof is valid for arbitrary state and symmetry of the interacting system.
      The finiteness of this potential energy
     at the nucleus in these local effective potential energy theories is a direct consequence of
     the electron-nucleus cusp condition. Since the cusp condition holds for both ground and excited states of the interacting system, the effective electron-interaction potential energy is finite at the nucleus whether the noninteracting fermions are in a ground or excited state. The proof is general, and does not distinguish between ground and excited states of the model system. The noninteracting bosons are, of course, always in their ground state.\\

     The study of the structure of these effective potential energies via densities derived
     from accurate wave functions now has an important proviso. These wave functions must
     satisfy the electron-nucleus cusp condition. Otherwise, the potential energies will
     be singular at a nucleus, thereby leading to erroneous conclusions regarding their
     structure.\\

\subsection{}
\subsubsection{}

\begin{acknowledgments}
\textbf{ACKNOWLEDGMENT}\\
This work was supported in part by the Research Foundation of
 CUNY. We thank Prof. A. J. Thakkar for providing us with the
 Hydrogen molecule wave function.

\end{acknowledgments}


\begin{references}
\bibitem{1} T. Kato, Commun. Pure Appl. Math, \textbf{10}, 151 (1957).
\bibitem{2} W. A. Bingel, Z. Naturforschg.  \textbf{18a},1249 (1963); Theoret. Chim. Acta (Berl) \textbf{8},54(1967).
\bibitem{3} R. T. Pack and W. Byers Brown, J. Chem. Phys. \textbf{45}, 556 (1966);
\bibitem{4}W. Kohn and L. J. Sham, Phys. Rev. \textbf{140}, A 1133 (1965);
 R.M. Dreizler and E. K. U. Gross, \textit{Density Functional Theory},(Springer-Verlag,1990);
  R. G. Parr and W. Yang, \textit{Density-Functional Theory of Atoms and Molecules}, (Oxford University Press, New York,1989).
\bibitem{5} R. T. Sharp and G. K. Horton, Phys. Rev. \textbf{30}, 317 (1953);
J. D. Talman and W. F. Shadwick, Phys. Rev. A \textbf{14}, 36(1976);
J. B. Krieger, Y. Li, and G. J. Iafrate, in \textit{Density Functional Theory}, edited by
E. K. U. Gross and R. M. Dreizler, NATO ASI Series, Series B: Physics, Vol 337 (Plenum, New York, 1995).

\bibitem{6} V. Sahni, Phys. Rev. A \textbf{55}, 1846, (1997); Top Curr chem. \textbf{182}, 1, (1996); Z. Qian and V. Sahni, Phys. Rev. A \textbf{57}, 2527 (1998);
Phys. Rev. B  \textbf{62}, 16364 (2000); V. Sahni, L. Massa, R. Singh, and M. Slamet, Phys.
Rev. Lett. \textbf{87}, 113002(2001); M. Slamet and V. Sahni, Int. J. Quantum Chem. \textbf{85}, 436(2001).
\bibitem{7} J. P. Perdew, R. G. Parr, M. Levy and J. L. Balduz, Phys. Rev. Lett. \textbf{49},
1691(1982).
\bibitem{8} M. Levy, J. P. Perdew, and V. Sahni, Phys. Rev. A \textbf{30}, 2745(1984).
\bibitem{9} C. O. Almbladh and U. von Barth, Phys. Rev. B \textbf{31}, 3231(1985).
\bibitem{10} Y. Wang and  R. G. Parr, Phys. Rev. A \textbf{47},
R1591(1993); R.Van Leeuwan and E. J. Baerends, Phys. Rev. A
\textbf{49}, 2421(1994).

\bibitem{11} D. W. Smith, S. Jagannathan, G. S. Handler, Int. J. Quantum Chem.
 Symp. \textbf{13}, 103(1979);  S. Jagannathan, Ph. D.Thesis, University of Georgia, 1979.
\bibitem{12} E. R. Davidson, Int. J. Quantum Chem. \textbf{37}, 811(1990).
\bibitem{13}  C. J. Umrigar and X. Gonze, Phys. Rev. A \textbf{50}, 3827(1994).
\bibitem{14} C. O. Almbladh, A. C. Pedroza, Phys. Rev. A \textbf{29}, 2322 (1984);
            A. C. Pedroza, Phys. Rev. A \textbf{33}, 804 (1986).
\bibitem{15} Q. Zhao, R. C. Morrison, R. G. Parr, Phys. Rev. A \textbf{50}, 2138(1994);
            R. C. Morrison, Q. Zhao,   Phys. Rev. A \textbf{51}, 1980(1995);
            C. Filippi, X. Gonze, C. J. Umrigar, in \textit{Recent Developments
            and Applications of Density Functional Theory}; J. Seminaro Ed. , Elsevier:
             Amsterdam, 1996.
\bibitem{16} J. Chen, R. O. Esquivel, and M. J. Stott, Philos. Mag. B \textbf{69}, 1001(1994).

\bibitem{17} O. V. Gristsenko, R. V. Leeuwen, E. J. Baerends, Phys. Rev. A \textbf{52},
1870(1995).
\bibitem{18} S. Liu, R. G. Parr, and A. Nagy, Phys. Rev. A \textbf{52}, 2645(1995).
\bibitem{19} R. Singh, L. Massa, and V. Sahni, Phys. Rev. A \textbf{60},
 4135(1999); D. C. Langreth and M. J. Mehl, Phys. Rev. B
\textbf{28}, 1809(1983); J. P. Perdew and Y. Wang, Phys. Rev. B
\textbf{33}, 8800(1986); J. P. Perdew, in \emph{Electronic
Structure of Solids' 91}, edited by P. Ziesche and H. Eschrig
(Akademic Verlag, Berlin,1991); A. P. Becke, Phys. Rev. A
\textbf{38}, 3098(1988).
\bibitem{20} Z. Qian and V. Sahni, Int. J. Quantum  Chem. \textbf{79}, 205(2000).
\bibitem{21} V. Sahni (unpublished).
\bibitem{22} J. Komasa and A. J. Thakkar, Mol. Phys. \textbf{78}, 1039(1993);
  Phys. Rev. A \textbf{49}, 965(1994).
\bibitem{23} W. Kolos, K. Szalewicz, and H. K. Monkhorst, J. Chem. Phys. \textbf{84}, 3278(1986).
\bibitem{24} W. Kolos and C. C. J. Roothaan, Rev. Mod. Phys.
\textbf{32}, 219(1960).





\end{references}

\end{document}